\documentclass[twocolumn,prl,amsmath,amssymb,superscriptaddress,showpacs]{revtex4}

\usepackage{graphicx}
\usepackage{dcolumn}
\usepackage{bm}

\def\reff#1{(\ref{#1})}

\newcommand{\ten}[1]{\overline{\overline{#1}}}

\newcommand {\vB}{{\bf B}}
\newcommand {\vH}{{\bf H}}
\newcommand {\vE}{{\bf E}}
\newcommand {\vD}{{\bf D}}

\newcommand {\vM}{{\bf M}}
\newcommand {\vP}{{\bf P}}

\newcommand {\vp}{{\bf p}}
\newcommand {\vm}{{\bf m}}
\newcommand {\vk}{{\bf k}}

\begin{document}


\title{Negative refraction in (bi)-isotropic periodic arrangements of chiral SRRs}

\author{L. Jelinek}
\email{l_jelinek@us.es} 
\affiliation{Departamento de Electr\'onica y Electromagnetismo, Universidad de Sevilla, 41012-Sevilla, Spain}

\author{R. Marqu\'es}
\email{marques@us.es} 
\affiliation{Departamento de Electr\'onica y Electromagnetismo, Universidad de Sevilla, 41012-Sevilla, Spain}

\author{F. Mesa}
\email{mesa@us.es}
\affiliation{Departamento de F\'isica Aplicada 1, Universidad de Sevilla, 41012-Sevilla, Spain}

\author{J. D. Baena}
\email{juan_dbd@us.es} 
\affiliation{Departamento de Electr\'onica y Electromagnetismo, Universidad de Sevilla, 41012-Sevilla, Spain}
\affiliation{Departamento de F\'isica Aplicada, Universidad Nacional de Colombia, Bogot\'a, Colombia}

\date{\today}

\begin{abstract}
Bi-isotropic and isotropic negative refractive index (NRI) 3D metamaterials made from
periodic arrangements of chiral split ring resonators (SRRs) are proposed and
demonstrated. An analytical theory for the characterization and design of these
metamaterials is provided and validated by careful full-wave electromagnetic simulations.
The reported results are expected to pave the way to the design of practical 3D
bi-isotropic and isotropic NRI metamaterials made from a single kind of inclusions. 
\end{abstract}


\maketitle

Most proposals for developing negative refractive index (NRI) metamaterials (e.g.
\cite{Smith-2000}, \cite{Eleftheriades-2002}, \cite{Vendik-2006}) make use of two kind of
inclusions, one of them to provide negative permittivity and the other to provide
negative permeability. On the other hand, artificial media made from metallic chiral
inclusions have been known since far by physicists \cite{Bose-1898}, \cite{Lindmann-1920}.
It is known that they present the most general linear and reciprocal constitutive
relations \cite{Kong} which, assuming 3D spatial isotropy and a time dependence of the
kind $\exp(-j\omega t)$, can be written as   
\begin{eqnarray}
    \vD &=& \varepsilon_0 (1+\chi_e) \vE + j \sqrt{\varepsilon_0\mu_0} \;
    \kappa \vH \label{D(EH)} \\ 
    \vB &=& -j \sqrt{\varepsilon_0\mu_0} \;\kappa\vE +
    \mu_0(1+\chi_m) \vH \label{B(EH)}\,, 
\end{eqnarray} 
where $\chi_e$, $\chi_m$ and $\kappa$ are the electric, magnetic and
cross-polarizabilities of the metamaterial which, for lossless media, are real quantities.
Media satisfying (\ref{D(EH)}) and (\ref{B(EH)}) are usually called \emph{bi-isotropc}, to
differentiate from usual isotropic media with $\kappa=0$. Since they present simultaneous
electric and magnetic properties, it has been guessed that they may open the way to NRI
metamaterials made of a single kind of inclusions. To the best of our knowledge, the first
proposal in this direction was made by S.Tretyakov in \cite{Tretyakov-2003}, and further
developed in \cite{Tretyakov-2005}. Also, in \cite{Pendry-Science-2004} a mixture of
metallic chiral inclusions and wires was proposed in order to obtain negative refraction.

Eigenwaves propagating through reciprocal bi-iso\-tro\-pic media are right- and
left-handed circulary polarized plane waves, whose dispersion relation is given by
\cite{Kong} 
\begin{equation}\label{k}
k^{\pm} = k_0\left(\sqrt{\mu_r\varepsilon_r} \pm\kappa\right) \,, 
\end{equation} 
where
$k_0=\omega\sqrt{\varepsilon_0\mu_0}$, and $\varepsilon_r=(1+\chi_e)$, $\mu_r=(1+\chi_m)$.
In order to reduce forbidden bands of propagation coming from complex
values of $k^\pm$, it is desirable that $\chi_e(\omega) \simeq \chi_m(\omega)$ so that
$\mu_r$ and $\epsilon_r$ always have the same sign. As explained in \cite{Tretyakov-2005}
and \cite{Marques-2006}, usually this condition also implies that 
\begin{equation}\label{balance}
\chi_e(\omega) \simeq \chi_m(\omega) \simeq |\kappa(\omega)| \,.
\end{equation} 
The conditions for negative refraction in transparent bi-isotropic
chiral media were investigated in \cite{Mackay-2005}. In lossless bi-isotropic media
satisfying the \emph{balance} condition \reff{balance}, these conditions reduces to
$\varepsilon_r < 0.5$ and $\mu_r < 0.5$ \cite{Marques-2006}, which is less restrictive than the
usual condition, $\varepsilon_r<0$ and $\mu_r<0$, for conventional isotropic medium
(i.e., a medium with $\kappa=0$). The price to pay for this enhanced bandwidth is that
only one of the eigenwaves of \reff{k} exhibits negative refraction, the other presenting
positive refraction \cite{Marques-2006}.  

Most previously proposed and studied bi-isotropic artificial media (see, for instance,
\cite{Tretyakov-2005}, \cite{Marques-2006} and references therein) were \emph{random}
arrangements of chiral inclusions. Random arrangements have the advantage of ensuring
isotropy provided the number of inclusions per unit volume is high enough. However,
multiple scattering in random arrangements of inclusions affects the effective
constitutive parameters of the mixture, leading to an imaginary part of the
susceptibilities to account for radiation losses \cite{Kong}. This important drawback of
radiation losses can be avoided by using \emph{periodic} arrangements of chiral
scatterers. Periodic arrangements also provide higher density of scatterers, which will
result in stronger electromagnetic responses and wider NRI frequency bands. In addition,
periodic arrangements can provide much better realiability and reproducibility than random
mixtures. Finally, wave propagation through periodic structures can be studied by means of
standard electromagnetic solvers, thus providing a validation method for the
homogenization procedure that has no counterpart in random structures. For all these
reasons periodic arrangements of chiral inclusions are  preferred over random
mixtures. However, care must be taken to obtain a truly isotropic periodic
metamaterial in order to keep the appropriate symmetries of the
crystal; otherwise couplings between elements may destroy isotropy even if the structure
appears as isotropic in the dipolar approximation \cite{Baena-2007}. In \cite{Baena-2007}
it has been shown that a cubic arrangement of SRRs satisfying the thetraedron or (in
Schoenflies notation) T-group of symmetry provides an useful basic design for this
purpose. Fig.1.a shows a cubic arrangement of chiral SRRs satisfying this symmetry. It is
a combination of six identical chiral SRRs, which are shown in detail in Fig.1.b. The
proposed chiral SRR is a modification of the original SRR design \cite{Pendry-1999} that
substitutes the edge-coupling between rings by a broad-side coupling
\cite{Marques-AP-2003}, also introducing some helicity. It should be noticed that the SRRs
on opposite sides of the cube of Fig.1.a are identical, which makes it possible to
develop a periodic metamaterial with a single cubic lattice from this
design. The unit cell of this metamaterial has periodicity $a$ (the edge of the cube)
and will contain three chiral SRRs.  Note that  the proposed chiral-SRRs is expected
to be a feasible ``practical'' design because of its possible manufacture via standard
planar-circuit fabrication techniques\footnote{In practical
designs a very low dielectric substrate (foam, for instance) could be introduced between
the rings in order to provide mechanical stability, and the bridges between them
substituted by standard via-hole connections}.
\begin{figure}
\centering
\includegraphics[width=0.8\columnwidth]{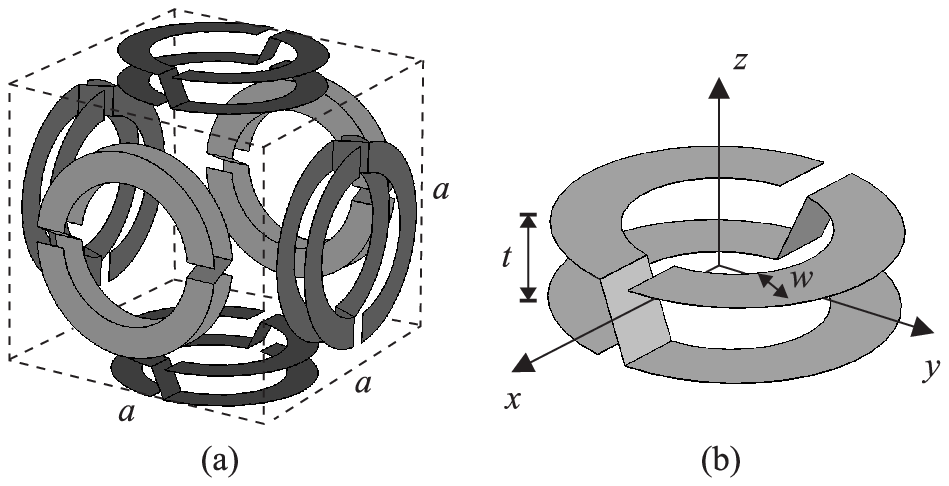}
\caption{\label{Fig1} (a) Cubic arrangement of chiral SRRs satisfying T-group of symmetry
(namely, any of the inbody diagonals is a third order rotation symmetry axis), and (b)
detail of the chiral SRR design.} 
\end{figure}

The chiral SRR polarizabilities are defined by 
\begin{eqnarray}\label{m} 
\vp &=& \ten{\alpha}^{\,\text{ee}}\cdot\vE_\text{l} +
\ten{\alpha}^{\,\text{em}}\cdot\vB_\text{l} \\
\label{p} 
\vm &=& -(\ten{\alpha}^{\,\text{em}})^\text{T}\cdot\vE_\text{l} +
\ten{\alpha}^{\,\text{mm}}\cdot\vB _\text{l} \,,
\end{eqnarray}
where $(\cdot)^\text{T}$ means transpose, $\vp$ and $\vm$ are the electric and magnetic
dipoles induced in the chiral SRR, $\vE_\text{l}$ and $\vB_\text{l}$ are the electric and
magnetic local fields \emph{seen} by the SRR, and Onsager symmetries \cite{Landau-8} for
the magneto-electric susceptibilities $\ten{\alpha}^{\,\text{me}} =
-(\ten{\alpha}^{\,\text{em}})^\text{T}$ have been explicitly introduced. Analytical
expressions for these polarizabilities can be obtained by a straighforward extension of
the theory reported in \cite{Marques-AP-2003}. The results are \cite{Marques-Roma-2007} 
\begin{eqnarray}
     \label{alphamm}
     \alpha_{zz}^\text{mm} &=& \alpha_0^\text{mm}
    X(\omega)
    \;;\;\;\;\;\;\;\;\;\;\;\;\;\;\;\;\;\;
    \alpha_0^\text{mm} = \frac{\pi^2r^4}{L}
      \\
      \label{alphaem}
     \alpha_{zz}^\text{em} &=& = \pm \alpha_0^\text{em}\,
     \left(\frac{\omega_0}{\omega}\right)\,
     X(\omega)
     ;\;\;
    \alpha_0^\text{em} = j\frac{2\pi r^2
    t}{\omega_0L}
      \\
     \label{alphaee}
     \alpha_{zz}^\text{ee} &=& \alpha_0^\text{ee}
     \left(\frac{\omega_0}{\omega}\right)^2
      X(\omega)
      \;;\;\;\;\;\;\;\;\;
    \alpha_0^\text{ee} = \frac{(2t)^2}{\omega_0^2L} \\
    \alpha_{xx}^\text{ee} &=& \alpha_{yy}^\text{ee} = \alpha_0
    \;;\;\;\;\;\;\;\;\;\;\;\;\;\; \;\;\;\;\;\;
    \alpha_0 = \varepsilon_0\frac{16}{3}r_\text{ext} \label{alpha0}
\end{eqnarray}
where $r_\text{ext}$ is the external radius of the inclusion, $r=r_\text{ext}-w/2$ is the
average radius, and $X(\omega)$ is a common resonant factor given by
\begin{equation}\label{X}
X(\omega) = \frac{\omega^2}{\omega_0^2-\omega^2+j \omega R/L} \,. 
\end{equation}
The frequency of resonance $\omega_0=1/\sqrt{LC}$ of the inclusion, as well as the self
inductance, $L$, the capacitance, $C$, and the resitance, $R$, coincide with those of the
aforementioned broadside-coupled SRR \cite{Marques-AP-2003}. From
\reff{alphamm}--\reff{alpha0} the average magnetic-, electric- and cross-polarizability
per particle in the isotropic metamaterial are easily obtained as
$\langle\alpha_\text{m}\rangle =
\alpha_{zz}^\text{mm}/3$, $\langle\alpha_\text{e}\rangle = \alpha_{zz}^\text{ee}/3 +
\alpha_{xx}^\text{ee}/3 + \alpha_{yy}^\text{ee}/3$, and $\langle\alpha_\text{em}\rangle =
\alpha_{zz}^\text{em}/3$ respectively. From these expresions, the susceptibilities of the
metamaterial can be obtained from a straighforward extension of Lorentz local field
theory that takes into account the presence of cross-polarizabilities. This theory leads
to the following equations for the macroscopic polarization, $\vP$, and magnetization,
$\vM$, of the metamaterial:
\begin{align}
\label{P} 
\vP & = N
\left\{\langle\alpha_\text{e}\rangle\left(\vE+\frac{\vP}{3\varepsilon_0}\right) +
\mu_0\langle\alpha_\text{em}\rangle\left(\vH+\frac{\vM}{3}\right)\right\}  \\
\label{M} 
\vM &= N
\left\{\mu_0\langle\alpha_\text{m}\rangle\left(\vH+\frac{\vM}{3}\right) -
\langle\alpha_\text{em}\rangle\left(\vE+\frac{\vP}{3\varepsilon_0}\right)\right\}\, 
\end{align} 
where $N=3/a^3$ is the number of particles per unit volume, and $\vE$,
$\vH$ are the macroscopic fields. Using \reff{P}--\reff{M} the metamaterial
susceptibilities are directly obtained from their definitions
$\chi_e=\vE/(\varepsilon_0\vP)$ and $\chi_m=\vH/\vM$. The balance condition \reff{balance}
is translated to the chiral-SRR polarizabilities as $\langle\alpha_\text{m}\rangle =
|\langle\alpha_\text{em}\rangle|/c = \langle\alpha_\text{e}\rangle/c^2$ where $c$ is the
velocity of light. If the effect of the non-resonant polarizability \reff{alpha0} is
neglected, the above condition reduces to the following approximate balance condition:
\begin{equation} \label{bc}
       t\lambda_0 = (\pi r)^2\;,
\end{equation} 
where $\lambda_0$ is the wavelength at resonance.

\begin{figure}
\centering
\includegraphics[width=0.8\columnwidth]{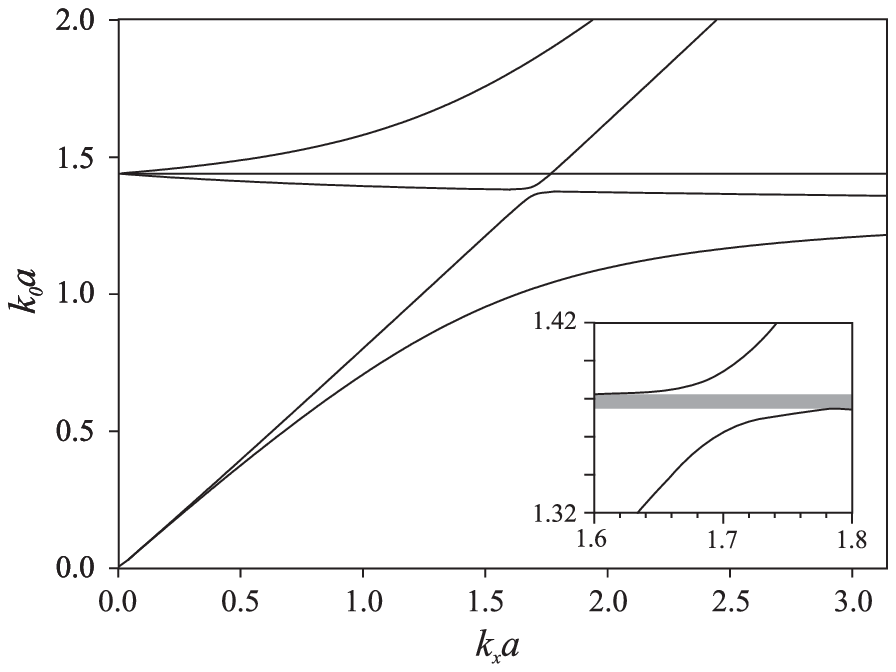}
\caption{\label{Fig2} Theoretical dispersion curves for the eigenwaves of the periodic
metamaterial made of the periodic repetition of the cubic arrangement of Fig.1.a, forming
a single cubic (sc) lattice with peridocity $a$. SRR parameters are $a/r_\text{ext}=2.87$,
$w/r_\text{ext}=0.5$, and $t/r_\text{ext}=0.47$. Dispersion curves are drawn inside the
first Brillouin zone of the structure.} 
\end{figure}

\begin{figure}
\centering
\includegraphics[width=0.8\columnwidth]{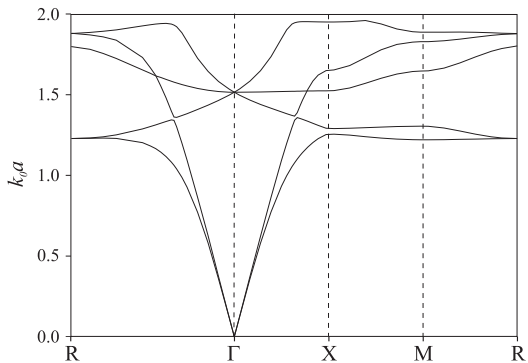}
\caption{\label{Fig3} Band structure and dispersion relation for the metamaterial of Fig.2
obtained from full-wave simulations along the path R-$\Gamma$-X-M-R.} 
\end{figure}

\begin{figure}
\centering
\includegraphics[width=0.8\columnwidth]{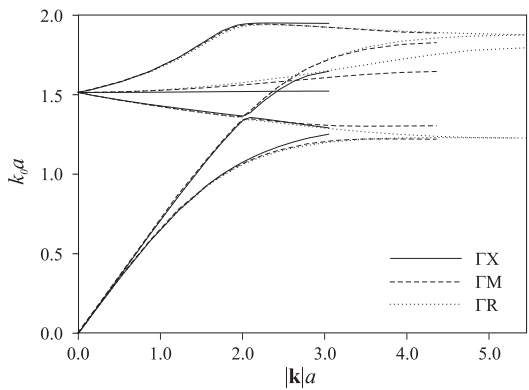}
\caption{\label{Fig4} Dispersion curves along the $\Gamma$-X, $\Gamma$-M, and  $\Gamma$-R
directions for the eigenwaves of  the balanced metamaterial of Figs. 2 and 3.} 
\end{figure}

\begin{figure}
\centering
\includegraphics[width=0.8\columnwidth]{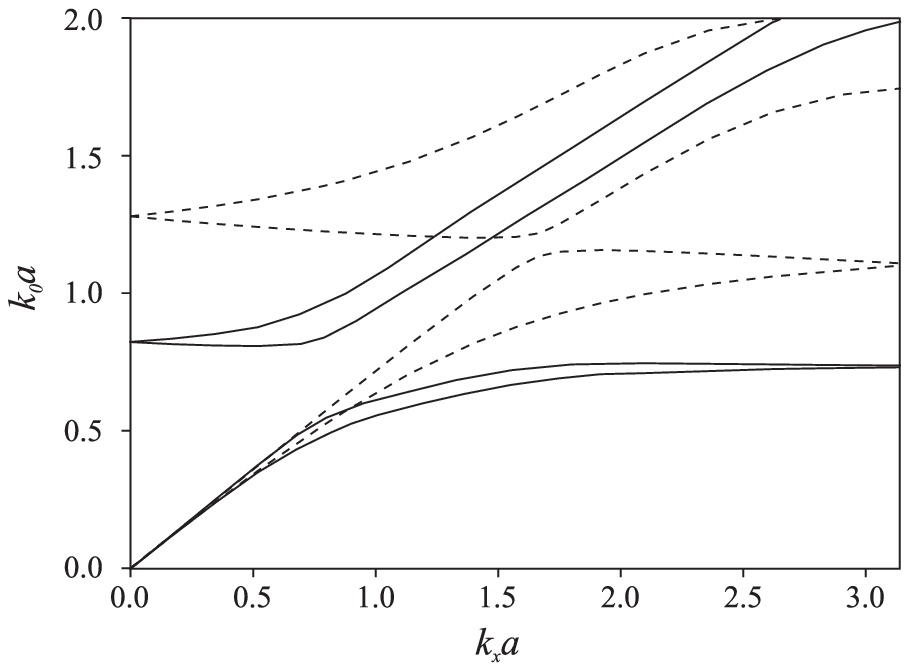}
\caption{\label{Fig5} Dispersion curves along the $\Gamma$-X direction for the transverse
eigenwaves of two unbalanced metamaterials. Parameters are tha same as in Figs.2 to 4,
except for $t/r_\text{ext}$, which is taken as $t/r_\text{ext}=0.3$ (dashed lines) and
$t/r_\text{ext}=0.1$ (solid lines)}  
\end{figure}

\begin{figure}
\centering
\includegraphics[width=0.8\columnwidth]{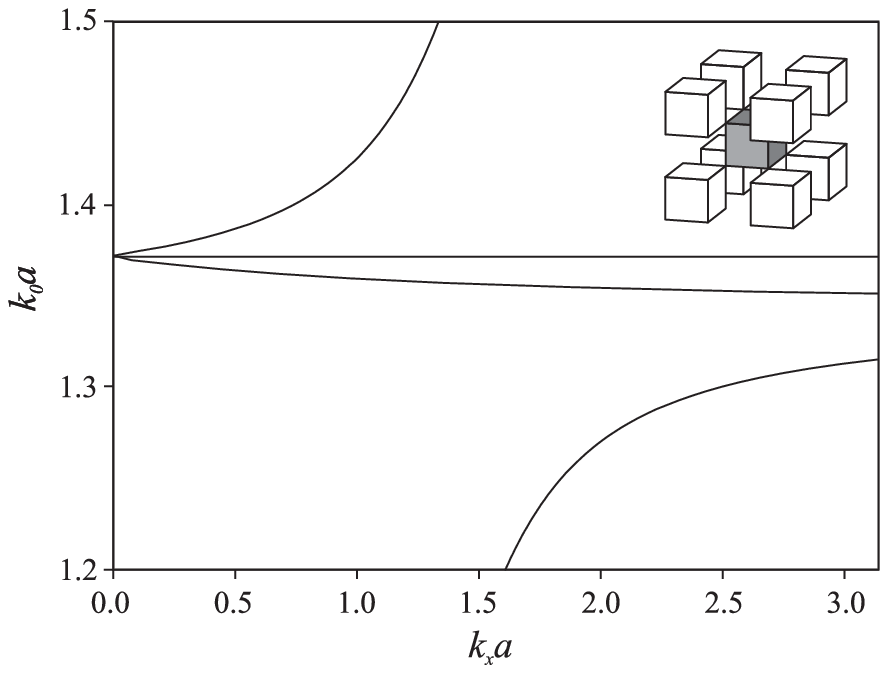}
\caption{\label{Fig6} Theoretical dispersion relation for plane waves propagating in the
racemic metamaterial with the single cubic lattice shown in the inset. Inset: Racemic
periodic structure made of two interleaved single cubic lattices of cubes as that of
Fig.1.a. White cubes are identical to that of Fig.1.a, and gray cubes are similar cubes
made of chiral SRRs of opposite handedness. Chiral SRR parameters are as in Fig.2 and
periodicity of the structure is $2a$. The fully degenerate longitudinal mode is now
located at $\varepsilon_r\mu_r=0$} 
\end{figure}

In the following a particular example is analyzed in order to illustrate and validate our
analytical theory. The chosen metamaterial  parameters satisfy Eq.~\ref{bc} and are given
in the caption of Fig.2. This figure shows the theoretical dispersion curves for the
eigenwaves of the bi-isotropic metamaterial. A region of backward-wave propagation
for one of the eigenwaves can be clearly observed. This frequency band corresponds to the
aforementioned condition $\varepsilon_r,\mu_r < 0.5$. Inside this band there is also a
small forbidden band gap (see the gray region in the inset of the figure) that
corresponds to complex values of the propagation constants \reff{k}. This small band gap
is due to the aforementioned
approximations implicit in Eq.~\ref{bc}. The straight horizontal line at the frequency
where the propagation constant of one of the eigenwaves vanishes (corresponding to the
condition $\varepsilon_r\mu_r = \kappa^2$) represents a fully degenerate longitudinal wave
with $\vk\times\vE=0$ and $\vk\times\vH=0$.  

The same structure has been simulated by using \emph{CST Microwave Studio} and its band
structure is shown in Fig.3. A good qualitative agreement is found between Figs.~2 and
3. In both figures a frequency band of backward-wave propagation is observed for one of
the eigenwaves. Also a small frequency stopband appears in Fig.3 inside this backward-wave
passband (because of the approximate balance conditition employed) in agreement with the
theoretical predictions of Fig.2. In order to show the
isotropy of the structure, the dispersion curves along the $\Gamma$-X, $\Gamma$-M and
$\Gamma$-R contours are shown in Fig.4. The isotropy of the dispersion relation for the
transverse modes becomes apparent from these curves, except for high values of the
propagation constant, where spatial dispersion should affect the dispersion relation. The
longitudinal mode of Fig.2 also appears in Figs.3 and 4, although now spatial dispersion
destroys degeneracy and isotropy. It is illustrative to compare the band structure of the reported
balanced structure with those of an unbalanced one. Thus, Fig.5 shows the dispersion
diagrams for the transverse modes of two unbalanced structres (which were obtained by
varying $t$ in Fig.1). It can be observed how, as $t$ decreases, the forbidden frequency
band increases until it occupies all the backward-wave region, giving a behavior similar
to that of a negative-$\mu$ split ring metamaterial \cite{Baena-2008}. This is expected,
since the electric and magneto-electric polarizabilities of the chiral-SRR
\reff{alphaem}-\reff{alphaee} decrease with $t$, until the chiral-SRR becomes an almost
purely magnetic inclusion. Fig.5 illustrates the tolerance of the proposed design to
deviations from the balance condition \reff{bc}. As it can be seen in the figure (dashed
lines) this tolerance is quite good. 

Let us now explore the possibility of designing a racemic mixture of chiral SRRs while
keeping the necessary symmetries to ensure an isotropic behavior for the metamaterial. For
this purpose let us consider a single cubic lattice of peridicity $2a$ made of cubes as
that shown in Fig.1.a, and another similar cubic lattice made of the same cubes but of
opposite handedness Both cubic lattices can be interleaved as shown in the inset of Fig.6
to give a single cubic lattice whose unit cell is formed by two cubes of opposite
handedness. Since the whole structure is symmetric after inversion (the centers of
symmetry are the corners of the cubes), the cross suceptibility $\kappa$ (which is a
pseudo-scalar) must vanish \cite{Landau-8}. Therefore the whole structure is a balanced isotropic
metamaterial with $\kappa=0$. Fig.6 shows the theoretical dispersion relation
($k=k_0\sqrt{\varepsilon_r\mu_r}$) of the proposed structure. A frequency band of
backward-wave propagation, corresponding to the condition $\varepsilon_r,\mu_r<0$, can be
observed in the figure.  

When designing a NRI metamaterial it is not sufficient to obtain some amount of negative
refraction but it is also required to have small reflectance so that the refracted beam
has a significant amplitude. As it is well known, in isotropic media the reflectance is
governed by the wave impedance $Z=Z_0\sqrt{\mu_r/\varepsilon_r}$, where $Z_0 =
\sqrt{\mu_0/\varepsilon_0} \approx 377 \,\Omega$ is the wave impedance of free space. For
bi-isotropic media the same expression $Z=Z_0\sqrt{\mu_r/\varepsilon_r}$ holds
\cite{Kong}. Since for balanced metamaterials it is imposed that $\chi_e\simeq\chi_m$ in
the frequency range of interest, it directly follows that $Z(\omega)\simeq Z_0(\omega)$,
which ensures good matching to free space. This last feature is an additional
relevant advantage of the proposed \emph{balanced} bi-isotropic and isotropic NRI
metamaterials. 

In summary, periodic NRI metamaterials based on chiral SRRs have been proposed and
demonstrated. An analytical theory has been developed for the characterization of such
metamaterias, which has been validated by careful full wave electromagnetic simulations.
It has been shown that single cubic lattices of chiral-SRRs can provide bi-isotropic NRI
metamaterials with a well defined frequency band of backward-wave propagation for one of
its plane-wave eigenstates. Besides, a racemic single cubic lattice of chiral SRRs
satisfying the apropriate symmetry (T$_h$ group in Schoenflies notation) has been proposed
in order to provide NRI isotropic metamaterials whith a well defined band of backward wave
propagation for all its plane-wave eigenstates. In both cases good matching to free space
is expected at all frequencies.  

\begin{acknowledgments}
This work has been supported by the Spanish Ministerio de Educaci\'on y Ciencia under
projects TEC2007-68013-C02-01/TCM and TEC2007-65376/TCM, and by Spanish Junta de Andaluc\'ia under projects
P06-TIC-01368 and TIC-253.
\end{acknowledgments}


\begin{thebibliography}{17}
\expandafter\ifx\csname natexlab\endcsname\relax\def\natexlab#1{#1}\fi
\expandafter\ifx\csname bibnamefont\endcsname\relax
  \def\bibnamefont#1{#1}\fi
\expandafter\ifx\csname bibfnamefont\endcsname\relax
  \def\bibfnamefont#1{#1}\fi
\expandafter\ifx\csname citenamefont\endcsname\relax
  \def\citenamefont#1{#1}\fi
\expandafter\ifx\csname url\endcsname\relax
  \def\url#1{\texttt{#1}}\fi
\expandafter\ifx\csname urlprefix\endcsname\relax\def\urlprefix{URL }\fi
\providecommand{\bibinfo}[2]{#2}
\providecommand{\eprint}[2][]{\url{#2}}

\bibitem[{\citenamefont{Smith et~al.}(2000)\citenamefont{Smith, Padilla, Vier,
  Nemat-Nasser, and Schultz}}]{Smith-2000}
\bibinfo{author}{\bibfnamefont{D.~R.} \bibnamefont{Smith}},
  \bibinfo{author}{\bibfnamefont{W.~J.} \bibnamefont{Padilla}},
  \bibinfo{author}{\bibfnamefont{D.~C.} \bibnamefont{Vier}},
  \bibinfo{author}{\bibfnamefont{S.~C.} \bibnamefont{Nemat-Nasser}},
  \bibnamefont{and} \bibinfo{author}{\bibfnamefont{S.}~\bibnamefont{Schultz}},
  \bibinfo{journal}{Phys. Rev. Lett.} \textbf{\bibinfo{volume}{84}},
  \bibinfo{pages}{4184} (\bibinfo{year}{2000}).

\bibitem[{\citenamefont{Eleftheriades et~al.}(2002)\citenamefont{Eleftheriades,
  Iyer, and Kremer}}]{Eleftheriades-2002}
\bibinfo{author}{\bibfnamefont{G.~V.} \bibnamefont{Eleftheriades}},
  \bibinfo{author}{\bibfnamefont{A.~K.} \bibnamefont{Iyer}}, \bibnamefont{and}
  \bibinfo{author}{\bibfnamefont{P.~C.} \bibnamefont{Kremer}},
  \bibinfo{journal}{IEEE Trans. on Microwave Theory and Tech.}
  \textbf{\bibinfo{volume}{50}}, \bibinfo{pages}{2702} (\bibinfo{year}{2002}).

\bibitem[{\citenamefont{Vendik et~al.}(2006)\citenamefont{Vendik, Vendik,
  Kolmarov, and Odit}}]{Vendik-2006}
\bibinfo{author}{\bibfnamefont{I.}~\bibnamefont{Vendik}},
  \bibinfo{author}{\bibfnamefont{O.}~\bibnamefont{Vendik}},
  \bibinfo{author}{\bibfnamefont{I.}~\bibnamefont{Kolmarov}}, \bibnamefont{and}
  \bibinfo{author}{\bibfnamefont{M.}~\bibnamefont{Odit}},
  \bibinfo{journal}{Opto-Electronics Rev.} \textbf{\bibinfo{volume}{14}},
  \bibinfo{pages}{179} (\bibinfo{year}{2006}).

\bibitem[{\citenamefont{Bose}(1898)}]{Bose-1898}
\bibinfo{author}{\bibfnamefont{J.~C.} \bibnamefont{Bose}},
  \bibinfo{journal}{Proc. R. Soc. Lond.} \textbf{\bibinfo{volume}{63}},
  \bibinfo{pages}{146} (\bibinfo{year}{1898}).

\bibitem[{\citenamefont{Lindmann}(1920)}]{Lindmann-1920}
\bibinfo{author}{\bibfnamefont{K.~F.} \bibnamefont{Lindmann}},
  \bibinfo{journal}{Annalen der Physik} \textbf{\bibinfo{volume}{63}},
  \bibinfo{pages}{621} (\bibinfo{year}{1920}).

\bibitem[{\citenamefont{Kong}(2000)}]{Kong}
\bibinfo{author}{\bibfnamefont{J.~A.} \bibnamefont{Kong}},
  \emph{\bibinfo{title}{Electromagnetic Wave Theory}} (\bibinfo{publisher}{EMW
  Publishing}, \bibinfo{year}{2000}).

\bibitem[{\citenamefont{Tretyakov}(2003)}]{Tretyakov-2003}
\bibinfo{author}{\bibfnamefont{S.}~\bibnamefont{Tretyakov}},
  \emph{\bibinfo{title}{Analytical Modeling in Applied Electromagnetism}}
  (\bibinfo{publisher}{Artech House.}, \bibinfo{year}{2003}).

\bibitem[{\citenamefont{Tretyakov et~al.}(2005)\citenamefont{Tretyakov,
  Sihvola, and Jylh}}]{Tretyakov-2005}
\bibinfo{author}{\bibfnamefont{S.~A.} \bibnamefont{Tretyakov}},
  \bibinfo{author}{\bibfnamefont{A.}~\bibnamefont{Sihvola}}, \bibnamefont{and}
  \bibinfo{author}{\bibfnamefont{L.}~\bibnamefont{Jylh}},
  \bibinfo{journal}{Photonics and Nanostruct. Fund. and Appl.}
  \textbf{\bibinfo{volume}{3}}, \bibinfo{pages}{107} (\bibinfo{year}{2005}).

\bibitem[{\citenamefont{Pendry}(2004)}]{Pendry-Science-2004}
\bibinfo{author}{\bibfnamefont{J.~B.} \bibnamefont{Pendry}},
  \bibinfo{journal}{Science} \textbf{\bibinfo{volume}{306}},
  \bibinfo{pages}{1353} (\bibinfo{year}{2004}).

\bibitem[{\citenamefont{Marqu\'es et~al.}(2006)\citenamefont{Marqu\'es,
  Jelinek, and Mesa}}]{Marques-2006}
\bibinfo{author}{\bibfnamefont{R.}~\bibnamefont{Marqu\'es}},
  \bibinfo{author}{\bibfnamefont{L.}~\bibnamefont{Jelinek}}, \bibnamefont{and}
  \bibinfo{author}{\bibfnamefont{F.}~\bibnamefont{Mesa}},
  \bibinfo{journal}{Micr. and Opt. Techn. Lett.} \textbf{\bibinfo{volume}{49}},
  \bibinfo{pages}{2006} (\bibinfo{year}{2006}).

\bibitem[{\citenamefont{Mackay}(2005)}]{Mackay-2005}
\bibinfo{author}{\bibfnamefont{T.~G.} \bibnamefont{Mackay}},
  \bibinfo{journal}{Microwave and Opt. Tech. Lett.}
  \textbf{\bibinfo{volume}{45}}, \bibinfo{pages}{120} (\bibinfo{year}{2005}).

\bibitem[{\citenamefont{Baena et~al.}(2007{\natexlab{a}})\citenamefont{Baena,
  Jelinek, and Marqu\'es}}]{Baena-2007}
\bibinfo{author}{\bibfnamefont{J.~D.} \bibnamefont{Baena}},
  \bibinfo{author}{\bibfnamefont{L.}~\bibnamefont{Jelinek}}, \bibnamefont{and}
  \bibinfo{author}{\bibfnamefont{R.}~\bibnamefont{Marqu\'es}},
  \bibinfo{journal}{Phys. Rev. B} \textbf{\bibinfo{volume}{76}},
  \bibinfo{pages}{245115} (\bibinfo{year}{2007}{\natexlab{a}}).

\bibitem[{\citenamefont{Pendry et~al.}(1999)\citenamefont{Pendry, Holden,
  Robbins, and Stewart}}]{Pendry-1999}
\bibinfo{author}{\bibfnamefont{J.~B.} \bibnamefont{Pendry}},
  \bibinfo{author}{\bibfnamefont{A.~J.} \bibnamefont{Holden}},
  \bibinfo{author}{\bibfnamefont{D.~J.} \bibnamefont{Robbins}},
  \bibnamefont{and} \bibinfo{author}{\bibfnamefont{W.~J.}
  \bibnamefont{Stewart}}, \bibinfo{journal}{IEEE Transactions Microwave Theory
  Tech.} \textbf{\bibinfo{volume}{47}}, \bibinfo{pages}{2075}
  (\bibinfo{year}{1999}).

\bibitem[{\citenamefont{Marqu\'es et~al.}(2003)\citenamefont{Marqu\'es, Mesa,
  Martel, and Medina}}]{Marques-AP-2003}
\bibinfo{author}{\bibfnamefont{R.}~\bibnamefont{Marqu\'es}},
  \bibinfo{author}{\bibfnamefont{F.}~\bibnamefont{Mesa}},
  \bibinfo{author}{\bibfnamefont{J.}~\bibnamefont{Martel}}, \bibnamefont{and}
  \bibinfo{author}{\bibfnamefont{F.}~\bibnamefont{Medina}},
  \bibinfo{journal}{IEEE Trans. on Antenas and Propagation}
  \textbf{\bibinfo{volume}{51}}, \bibinfo{pages}{2572} (\bibinfo{year}{2003}).

\bibitem[{\citenamefont{Landau et~al.}(1984)\citenamefont{Landau, Lifshitz, and
  Pitaevskii}}]{Landau-8}
\bibinfo{author}{\bibfnamefont{L.~D.} \bibnamefont{Landau}},
  \bibinfo{author}{\bibfnamefont{E.~M.} \bibnamefont{Lifshitz}},
  \bibnamefont{and} \bibinfo{author}{\bibfnamefont{L.~P.}
  \bibnamefont{Pitaevskii}}, \emph{\bibinfo{title}{Electrodynamics of
  Continuous Media}} (\bibinfo{publisher}{Pergamon}, \bibinfo{year}{1984}),
  \bibinfo{edition}{3rd} ed.

\bibitem[{\citenamefont{Marqu\'es et~al.}(2007)\citenamefont{Marqu\'es, Mesa,
  Jelinek, and Baena}}]{Marques-Roma-2007}
\bibinfo{author}{\bibfnamefont{R.}~\bibnamefont{Marqu\'es}},
  \bibinfo{author}{\bibfnamefont{F.}~\bibnamefont{Mesa}},
  \bibinfo{author}{\bibfnamefont{L.}~\bibnamefont{Jelinek}}, \bibnamefont{and}
  \bibinfo{author}{\bibfnamefont{J.~D.} \bibnamefont{Baena}}, in
  \emph{\bibinfo{booktitle}{Proc. Metamaterials 2007}} (\bibinfo{address}{Rome
  (Italy)}, \bibinfo{year}{2007}), p. \bibinfo{pages}{214}.

\bibitem[{\citenamefont{Baena et~al.}(2007{\natexlab{b}})\citenamefont{Baena,
  Jelinek, and Marqu\'es}}]{Baena-2008}
\bibinfo{author}{\bibfnamefont{J.~D.} \bibnamefont{Baena}},
  \bibinfo{author}{\bibfnamefont{L.}~\bibnamefont{Jelinek}}, \bibnamefont{and}
  \bibinfo{author}{\bibfnamefont{R.}~\bibnamefont{Marqu\'es}},
  \bibinfo{journal}{http://arxiv.org/abs/0711.4215}
  (\bibinfo{year}{2007}{\natexlab{b}}).

\end{thebibliography}
\end{document}